\documentclass[preprint,
]{revtex4}
\newcommand{\be}{\begin{equation}}
\newcommand{\ee}{\end{equation}}
\newcommand{\bea}{\begin{eqnarray}}
\newcommand{\eea}{\end{eqnarray}}
\usepackage{graphicx,color}
\usepackage{epsfig}
\usepackage{bm}
\usepackage{float}

\usepackage{amssymb}

\begin{document}
\title{Markovian Description of Unbiased Polymer Translocation}
\author{Felipe Mondaini$^{1,2}$}
\author{L. Moriconi$^1$}
\affiliation{$^1$Instituto de F\'\i sica, Universidade Federal do Rio de Janeiro, \\
C.P. 68528, 21945-970, Rio de Janeiro, RJ, Brazil \\
$^2$Centro Federal de Educa\c c\~ao Tecnol\'ogica Celso Suckow da Fonseca, \\
UnED Angra dos Reis, Angra dos Reis, 23953-030, RJ, Brazil}
\begin{abstract}
We perform, with the help of cloud computing resources, extensive Langevin
simulations which provide compelling evidence in favor of a general
markovian framework for unbiased polymer translocation. Our statistical
analysis consists of careful evaluations of (i) two-point correlation functions
of the translocation coordinate and (ii) the empirical probabilities
of complete polymer translocation (taken as a function of the initial
number of monomers on a given side of the membrane). We find good agreement
with predictions derived from the Markov chain approach recently addressed
in the literature by the present authors.
\end{abstract}
\maketitle

\section{Introduction}

The phenomenon of polymer translocation through membrane pores has been the subject of a flury of research activity in recent years~\cite{muthu}, a fact related to its relevance to the understanding and development of important biotechnological processes, like DNA sequencing, gene therapy and cytoplasmic drug delivery in living cells~\cite{luo,zanta,kasia,fologea}. Computer simulations have been playing a dominant role in most of the expressive literature of polymer translocation, which may be, roughly, classified into the topical issues of i) translocation driven by chemical potential gradients~\cite{kasia, grosberg, huopa, kaifu, wei,luo2}, ii) translocation driven by external forces~\cite{luo2, bhatta, panja_bark,iko_etal}, and iii) unbiased translocation~\cite{wei, luo2,gauthier, panja}. One is usually interested to compute (and, eventually, to model) the scaling exponents of the power laws which are found to relate the average translocation time $\tau$ to the polymer size $N$.

Unbiased translocation, where the diffusion of a polymer through a membrane pore occurs uniquely as the consequence of thermal fluctuations, is by far the most studied case and it is also our focus in this paper. Chuang, Kantor and Kardar (CKK)~\cite{kardar} have introduced a successful description of unbiased homopolymer translocation from the simple assumption that the polymer's evolution does not take it far from its equilibrium states. The essential physical picture is that unbiased translocation is ultimately due to the diffusion of the polymer center of mass. The translocation time is, thus, assumed to scale precisely in the same way as the Rouse relaxation time~\cite{rouse}, so that $\tau \sim N^{1+2\nu}$, where $\nu$ is the well-known Flory exponent ($\nu \simeq 0.588$ in three-dimensions).

Since the scaling behavior $\tau \sim N^{1+2\nu}$ departs from the one of usual brownian diffusion, $\tau \propto N^2$, it has been eventually suggested~\cite{panja2,dubbeldam} that unbiased polymer translocation could not be modeled as a markovian process -- in other words, memory effects should be taken into account as an essential ingredient in any kinematical description of polymer translocation. However, it is of crucial importance to note that particular values of the translocation exponent are not sufficient {\it{per se}} to rule out the markovian nature of polymer translocation. A Markov chain approach can be put forward which actually leads to the CKK translocation exponent and to a closed analytical expression for the probability of complete polymer translocation that stands in good agreement with results obtained from Langevin simulations~\cite{Mondaini_Moriconi}.

Our central aim in this work is to subject the markovian hypothesis of unbiased polymer translocation to a more stringent test. In essential words, we have considered statistical ensembles of three-dimensional polymer translocation realizations, taken from Langevin simulations, with sizes considerably larger than the ones presently found in the literature. Our statistical data sets have been produced within the Grid Initiatives for e-Science virtual communities in Europe and Latin America (GISELA), a cloud computing framework supported by several academic institutions.

This paper is organized as follows. In section II, we discuss the Langevin dynamics of translocating polymers, which are modeled as bead-spring chains of Lennard-Jones particles with the finite-extension nonlinear elastic (FENE) potential. As a starting point, our simulations are validated from evaluations of the mean translocation time, which is verified to be compatible with the CKK scaling prediction~\cite{kardar}. We, then, compute two-point correlators of the translocation coordinate and, from them, the memory time of monomer translocation events. The polymer diffusion through the membrane pore is noted to be free of anomalies for times larger than the memory time scale, clearly suggesting that unbiased polymer translocation is essentially Markovian. Once we have found, actually, that small monomer clusters are correlated during translocation, we take this information into account in section III to compare, with good agreement, the empirical probabilities of complete translocation for polymers of various sizes to the theoretical probabilities obtained from the Markov chain approach of Ref.~\cite{Mondaini_Moriconi}. In section IV, we summarize our results and point out directions of further research.

\section{Langevin Simulations}

In our Langevin simulations, the excluded-volume and van der Waals interactions between beads (monomers and membrane atoms) separated by a distance $r$ are modeled through a repulsive Lennard-Jones (LJ) potential with cutoff at length $2^{1/6}\sigma$, where $\sigma$ is the bead diameter:
\be
U_{LJ}(r) = \left\{ \begin{array}{ll}
4 \epsilon [(\sigma/r)^{12}-(\sigma/r)^6]+\epsilon  \ , \
{\hbox{ if }} r \leq 2^{1/6} \sigma \ , \ \\
0  \ , \  {\hbox{ if }} r > 2^{1/6} \sigma \ . \
\end{array}\right.
\ee
Besides the LJ potential, consecutive monomers are subjet to the
Finite-Extension Nonlinear Elastic (FENE) potential,
\be
U_{F}(r) = - \frac{1}{2} k R_0^2 \ln [ 1 -(r/R_0)^2] \ . \
\ee
From the above definition, it is clear that the FENE potential does not allow the distance between consecutive monomers to become larger than $R_0$.

We have studied polymers with sizes up to 300 monomers, which translocate through a pore created by the remotion of a single atom at the center of an 80 x 80 monoatomic square lattice membrane. Translocation is dynamically described by the following Langevin equations,
\be
m \frac{d^2 \vec r_i}{dt} = - \sum_{ j \neq i} \vec \nabla_{r_i}  [U_{LJ}(r_{ij}) + U_F(r_{ij})] -\xi \frac{d \vec r_i }{dt} + \vec F_i(t) \ , \ \label{eq-motion}
\ee
where $r_{ij} = |\vec r_i - \vec r_j|$, $\xi$ is the dissipative constant and $\vec F_i(t)$ is a gaussian stochastic force which acts on the monomer with label $i$, completely defined from the expectation values
\bea
&&\langle \vec F_i(t) \rangle = 0 \ , \ \nonumber \\
&&\langle [\hat n \cdot \vec F_i(t)] [\hat n' \cdot \vec F_j(t') ] \rangle = 2 \hat n \cdot \hat n' k_B T \xi \delta_{ij} \delta(t-t') \ . \
\eea
Above, $\hat n$ and $\hat n'$ are arbitrary unit vectors, and $k_B$ and $T$ are the Boltzmann constant and the temperature, respectively. By means of a suitable regularization of the stochastic force, we have implemented a fourth-order Runge-Kutta scheme for the numerical simulation of the Langevin Equations (\ref{eq-motion}). Our simulation parameters are: $\epsilon=1.0$, $\sigma=1.0$ ($\sigma$ is also identified with the membrane lattice parameter), $\xi =0.7$, $k=7 \epsilon/\sigma^2$, $R_0 = 2 \sigma$, $k_B T = 1.2 \epsilon$. The simulation time step is taken to be $3 \times 10^{-3}t_{LJ}$, where $t_{LJ} \equiv \sqrt{m \sigma^2 / \epsilon}$ is the usual Lenard-Jones time scale. In the most general case, the initial configuration of the polymer has $n$ monomers on the trans-side of the membrane and $N-n$ on the cis-side. Translocation is allowed to start only after an initial stage of thermal equilibrium is reached for the whole polymer.

\begin{figure}[tbph]
\begin{center}
\includegraphics[width=11.88cm, height=9.24cm]{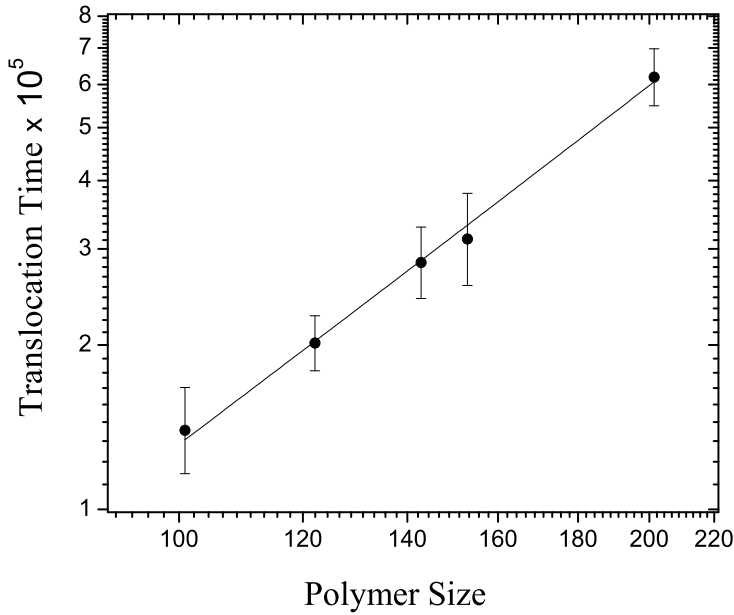}
\caption{The mean translocation time as a function of polymer size for unbiased translocation. We find the translocation exponent $\alpha = 2.17 \pm 0.06$, which is compatible with the CKK prediction, $\alpha = 1 + 2 \nu$, with $\nu=0.588$ (numerical value of the three-dimensional Flory exponent).}
\label{Tau}
\end{center}
\end{figure}

As a preliminary validation step, we have checked if the translocation samples produced from the Langevin simulations would lead, in fact, to mean translocation times that scale with the polymer size as expected on the grounds of the CKK phenomenological theory, i.e., $\tau \sim N^\alpha$ with $\alpha = 1+2 \nu$. The polymers are initially prepared to be in thermal equilibrium with an equal number of monomers on both sides of the membrane. Mean translocation times have been obtained from averages taken over ensembles of 70 complete translocation processes for each given polymer size. The mean translocation time as a function of the polymer size is shown in Fig. \ref{Tau}, which in fact indicates a reasonable agreement with the CKK scaling exponent. Being confident on the Langevin simulational scheme, we are now ready to move to the study of more subtle aspects of polymer translocation.
\vspace{0.2cm}

\leftline{\it{Correlation Effects}}
\vspace{0.2cm}

An important issue addressed in studies of polymer translocation refers to the role of correlations between monomer translocation events. As we show below, relying upon clear numerical evidence, such correlations are short-ranged in time, a fact that paves the way for a markovian description of unbiased translocation.
\begin{figure}[tbph]
\begin{center}
\includegraphics[width=11.88cm, height=9.24cm]{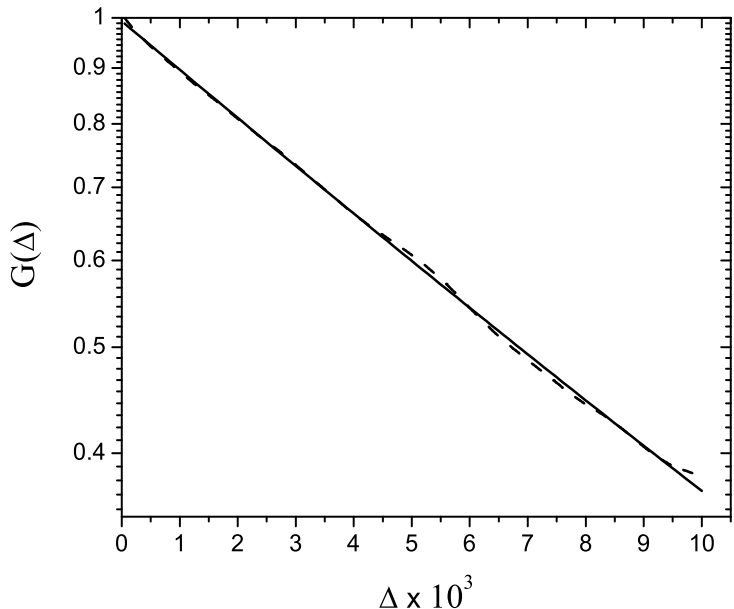}
\caption{The dashed line is the empirical evaluation of $G(\Delta)$ as given by
the truncated version of Eq. (\ref{c_funct}) with finite $M$ and $T$.
The solid line is the excellent fit provided by $\exp(-\Delta/9300)$.}
\label{CL}
\end{center}
\end{figure}

When a given dynamical system is claimed to evolve in time as a markovian stochastic process, it is of course tacitly assumed that the states of the system have been picked up in time intervals which are larger than some time scale associated to memory effects. In the polymer translocation context, it is convenient to define the dynamical state by the integer number $s(t)$, the ``translocation coordinate", that represents the number of monomers that have crossed the membrane to one of its sides, up to time $t$ (which, by the way, is also treated as a discrete variable, since we measure it in units of the Lennard-Jones time scale). The memory time scale can be defined, in principle, from the decaying profile of the normalized two-point correlation function
\be
G(\Delta) = \lim_{M,T \rightarrow \infty} \frac{\sum_{i=1}^M \sum_{t=0}^T s_i(t)s_i(t+\Delta)}{\sum_{i=1}^M \sum_{t=0}^T [s_i(t)]^2} \ , \ \label{c_funct}
\ee
where $s_i(t)$ denotes the $i$-th sample taken from the ensemble of translocation coordinate time series.
We have evaluated the right-hand side of (\ref{c_funct}) for polymers composed of $N=300$ monomers, in an ensemble of $M=1000$ samples, with time bound $T= 1.5 \times 10^4$. The initial polymer configurations are in thermal equilibrium with 150 monomers on each side of the membrane. As it is depicted in Fig. \ref{CL}, the two-point correlation function (\ref{c_funct}) follows, to very good approximation, the simple exponential law $G(\Delta) = \exp(-c \Delta)$, with decaying time parameter $1/c =9.3 \times 10^3$. In concrete terms, this result means that after around $9.3 \times 10^3$ Langevin iterations, translocation looses memory of the past states. It is interesting, having in mind Markov modeling, to find the typical size of correlated monomer clusters that translocate within the memory time scale. As we discuss below, this can be achieved from an analysis of the diffusive behavior of polymer translocation.
\vspace{0.2cm}

\leftline{\it{The normal diffusive regime of unbiased translocation}}
\vspace{0.2cm}

\begin{figure}[tbph]
\begin{center}
\includegraphics[width=11.88cm, height=9.24cm]{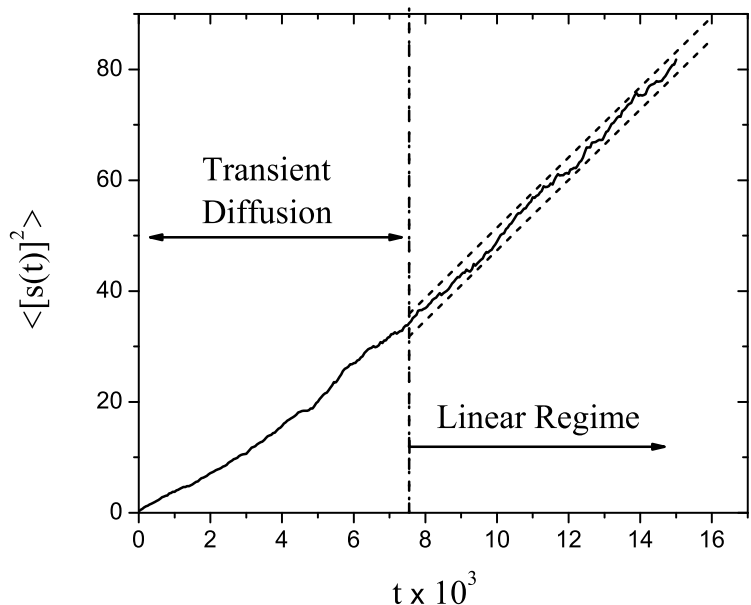}
\caption{Plot of the variance of the translocation coordinate as function of time for N=300. A linear regime effectively holds for $t \gtrsim 8 \times 10^3$, after an initial period of nonlinear transient behavior. Statistical averaging is performed in an ensemble of $10^4$ independent polymer translocation realizations.}
\label{s2linear}
\end{center}
\end{figure}

Still considering polymers of size $N=300$ in three dimensions, the variance of the translocation coordinate, that is, $\langle [s(t)]^2 \rangle$, is plotted in Fig. \ref{s2linear}. Averages are now taken over ensembles of $10^4$ samples.
It is clear from that picture that for times larger than the memory time scale, $\langle [s(t)]^2 \rangle$ is essentially a linear function of time or, in equivalent words, monomers diffuse in a normal way through the membrane pore.

It is important to emphasize that our data provides a strong objection to the previously predicted anomalous scaling
$\langle [s(t)]^2 \rangle \sim t^\frac{2}{1+2 \nu}$ ~\cite{kardar}, which must be viewed now as a misleading result
derived within the CKK theory of the mean translocation time. Fig. \ref{s2linear} also indicates that around the memory time scale $t=9.3 \times 10^3$ we have $\sqrt{\langle [s(t)]^2 \rangle} \simeq 6.5$, which means that the correlated monomer clusters are composed of approximately six monomers (we choose to round the size of the monomer clusters to six and not to seven, once we observe that the crossover to the linear regime in Fig. \ref{s2linear} takes place a little before the memory time scale set by the two-point correlation function (\ref{c_funct})). The size of such correlated monomer clusters is a crucial ingredient in the Markov chain approach to polymer translocation: if the original polymer is replaced by a ``monomer-clustered" polymer (whose size is the original size divided by the size of correlated monomer clusters), then the translocation of monomer clusters, rather than individual monomers, is assumed to generate a truly markovian stochastic process.

\section{Evidence of Markovian Behavior}

The fundamental hypothesis of the Markov chain approach to unbiased translocation is that monomers translocate in an uncorrelated way with probabilities $p_n$ and $q_n$ for cis$\rightarrow$trans and trans$\rightarrow$cis transitions, respectively (recall that $n$, as defined in Sec. II, is the number of monomers on the trans-side of the membrane). Following Ref.~\cite{Mondaini_Moriconi}, one puts forward the transition probabilities
\bea
&&p_n = \frac{c}{(N-n)^{\delta+2 \nu - 1}} \ , \ \nonumber \\
&&q_n = \frac{c} { n^{\delta+2 \nu - 1}} \ , \
\label{prob}
\eea
where $0<c<1$ is an arbitrary constant and $\delta$ is a scaling exponent associated to finite-size corrections to the CKK scaling relations (it would follow, for instance, that $\tau \sim N^{\delta + 2 \nu}$ for fixed $\delta$ and large enough $N$). It is also possible, from the definitions (\ref{prob}) and using exact results for general Markov chains~\cite{vanKampen}, to find the probability $P(N,n)$ of complete trans$\rightarrow$cis polymer translocation,
\be
P(N,n) = \frac{1 + \sum_{i=1}^{n-1} \prod_{j=1}^i \frac{q_j}{p_j}}{1 + \sum_{i=1}^{N-1} \prod_{j=1}^i \frac{q_j}{p_j}} \ . \
\label{pn}
\ee
\begin{figure}[tbph]
\begin{center}
\includegraphics[width=16.88cm, height=12.24cm]{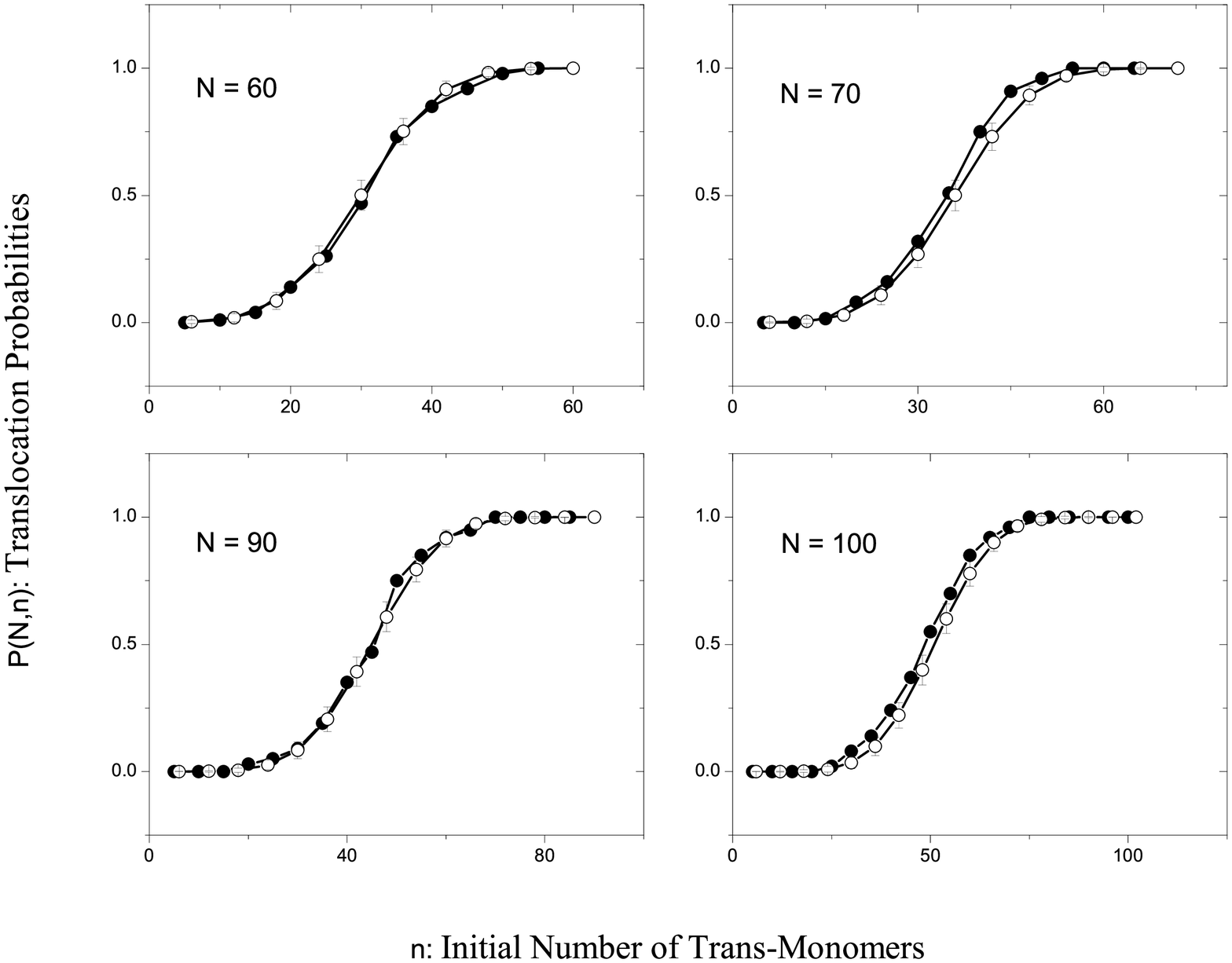}
\caption{The empirical (open circles) and analytical (solid circles) probabilities (both denoted here by $P(N,n)$) of cis $\rightarrow$ trans complete translocations are compared. The variable $n$ stands for the initial number of monomers in the trans-side of the membrane. The analytical probabilities are computed from (\ref{prob}) and (\ref{pn}) with $\delta=0.88$ and $\nu=0.588$ for $N=60,70,90,100$. The theoretical error bars follow from elementary statistical considerations and are given by $\sqrt{P(N,n)(1-P(N,n))/70}$.}
\label{Probability}
\end{center}
\end{figure}
We have tested the analytical prediction given by Eq. (\ref{pn}) for polymers of various sizes. As it is shown in Fig. \ref{Probability}, the comparison between the empirical and analytical probabilities is very satisfactory. The empirical probabilities have been evaluated from ensembles of 70 complete polymer translocation realizations. To understand Fig. \ref{Probability}, note that the parameter $N$ to be substituted in (\ref{prob}) is not the original polymer size anymore. In (\ref{prob}), $N$ is now taken as the effective size of the monomer-clustered polymer (in our particular case, as suggested by the results of Sec. II, it is just the original polymer size divided by 6). Therefore, for a given value $n$ of the initial number of trans-monomers, a solid circle is plotted in Fig. \ref{Probability}, with coordinates $(n, P(N/6,n/6))$ where $P(N/6,n/6)$ is the probability of complete polymer translocation evaluated within the Markov chain approach for a polymer which contains $N/6$ monomers ($N/6$ is conventionally rounded, if necessary to the smallest integer greater than $N/6$) .

\section{Conclusions}

We have provided consistent statistical data which essentially settles down the issue on whether unbiased polymer translocation is markovian or not -- it turns out that unbiased polymer translocation can be very confidently described as a markovian stochastic process. This conclusion is supported from three clear pieces of evidence: (i) the two-point correlation function (\ref{c_funct}) indicates that individual monomer translocation events are correlated whithin a finite memory time scale which is much smaller than the mean complete translocation time; (ii) the variance of the translocation coordinate depends linearly on time for times which are larger than the memory time scale; and (iii) empirical probabilities of complete translocation finely match the analytical ones predicted from the Markov chain approach of Ref.~\cite{Mondaini_Moriconi} (it is worth of mentioning that (ii) can be also derived within the same formalism).

An interesting point, motivated by our results, is whether the original approach of polymer translocation addressed by Muthukumar~\cite{muthu2}, which is also markovian (but failed to give the correct expression for the mean translocation time), can be somehow improved taking into account the present findings. In concrete terms, it is not difficult to define a specific free-energy profile for the translocating polymer that would lead to the individual monomer translocation probabilities (\ref{prob}). A challeging problem which we are currently investigating is whether such effective free-energy profile is just an artifact that can be used to reproduce the stochastic evolution of the translocating polymer, or is actually the physical thermodynamical potential derived from standard equilibrium statistical mechanics considerations.

As a final remark, we call attention to recent works on driven translocation, where memory effects are found to be relevant, due to the excitation of collective modes along the polymer chain \cite{panja_bark, iko_etal}. However, as a problem deserved for further research, we note that if the forces which pull the polymer are not strong enough, it is likely that translocation can still be described with the help of Markov chain equations, where the transition probabilities (\ref{prob}) are just replaced by alternative ones.
\vspace{0.3cm}

This work has been partially supported by CNPq and FAPERJ.

\end{document}